\documentclass{ab}

\usepackage{graphicx}
\usepackage{latexsym}

\def\degr{\hbox{$^\circ$}}

\begin{document}

\title{Reduction of CCD Observations Made with the Fabry-Perot Scanning Interferometer. II. Additional Procedures}

\author{A.V.~Moiseev$^1$ \and O.V.Egorov$^2$}
\institute{Special Astrophysical Observatory Russian Academy of Sciences,
N. Arkhyz, KChR, 369167, Russia \and
Sternberg Astronomical Institute, Universitetskii pr. 13, Moscow, 119992 Russia}
%
\date{November 12, 2007/Revised: November 20, 2007}
\offprints{A.V.  Moiseev, \email{moisav@sao.ru}}

\titlerunning{Reduction of CCD observations obtained with the scanning FPI}
\authorrunning{Moiseev, Egorov}%

\abstract{
We describe a software package used at the Special Astrophysical Observatory of the Russian
Academy of Sciences to reduce and analyze the data obtained with the Fabry-Perot scanning
interferometer. We already described most of the algorithms employed in our earlier
Paper I (Moiseev, \cite{mois02:Moiseev_n}). In this paper we focus on extra procedures required in the case
of the use of a high-resolution Fabry-Perot interferometer: removal of ghosts and
measurement of the velocity dispersion of ionized gas in galactic and extragalactic objects.
}

\maketitle

\section{Introduction}

Scanning Fabry-Perot interferometer (FPI) can be used to perform a detailed analysis of the
structure and kinematics of galaxies, nebulae, and other extended objects. Detailed
descriptions of the main idea of this observational method and references to earlier papers can found in Moiseev~(\cite{mois02:Moiseev_n}) (hereafter Paper~I) and Gordon~et~al.~(\cite{gordon00:Moiseev_n}). The result of observations has the form of a set of two-dimensional interferograms---the convolution of the monochromatic image of the object  with the transmission curve of the FPI at each step of scanning. After special reduction (phase-shift correction) these interferograms can be assembled into a
``data cube''. In such a cube, each spatial element in the detector plane is associated with its individual spectrum. The spectral interval is usually not very wide and amounts to only
5--50\,\AA, making it possible to study, e.g., the kinematics of ionized gas, based on the data for one or two emission lines.

To reduce observational data obtained with the FPI, one needs appropriate software, which
differs from the software employed to analyze the data obtained with slit spectrographs.
Paper~I gives a brief review of the systems commonly used to reduce
such data. Of the recently published papers on the subject we point out the paper by Daigle et
al. (\cite{daigle06:Moiseev_n}) who suggested a number of new procedures: adaptive spatial
binning of data cubes prior to constructing the velocity fields, algorithms for detecting emission lines and subtraction of night-sky lines.

At the Special Astrophysical Observatory of the Russian Academy of Sciences (SAO RAS) the
scanning FPI is a part of the CCD-based SCORPIO multimode focal reducer
(Afanasiev \&  Moiseev, \cite{afanas05:Moiseev_n}). Unlike
two-dimensional photon counters, which made a return recently and
are now used in combination with the FPI (Gach et al., \cite{gach02:Moiseev_n}), ``slower'' CCDs require a different kind of algorithms to reduce the data obtained. This concerns, first and foremost, photometric correction of channels and  night-sky spectrum subtraction. We addressed this issue in detail in our earlier Paper~I, where we described the algorithms used in the software to reduce SCORPIO observations performed in the
scanning FPI mode.

According to  ASPID\footnote{\tt http://alcor.sao.ru/db/aspid/} database, a total of about 180
FPI data cubes have been obtained with the 6-m telescope of SAO RAS in 2000--2007, with at
least 30 publications based on these data.
The experience in the reduction of this observational data allowed
us to improve our software package and develop a number of new useful procedures for data
analysis. In this paper we focus on describing these procedures. In section~\ref{ghost:Moiseev_n} we address the problem of removing the ghost light, which arises in high-resolution  FPIs,  in section~\ref{disp:Moiseev_n} we consider  the problem of velocity dispersion measurements and in section~\ref{soft:Moiseev_n} briefly describe the software package currently employed to  reduce observations made with the 6-m telescope using the scanning FPI incorporated  into SCORPIO instrument.

\section{Ghosts subtraction}
\label{ghost:Moiseev_n}

SCORPIO is now equipped with two scanning interferometers
(hereafter referred to as IFP235 and IFP501) with a gap between
the plates corresponding to 235 and 501 orders of interference,
respectively, at the wavelength of  $\lambda 6562.8$\AA. See
Paper~I for detailed parameters of both interferometers.
Because of the FPI location in the output pupil of the system
between the collimator and the camera of the focal reducer ghost
lights appear on the image. Ghost lights are due to backlight
reflecting inside the interferometer plates and between these
plates and the nearest lenses of the focal reducer, surfaces of
the narrow-band filter, etc. This is a well known problem, and
various ghost ``families'' were described by
Bland-Hawthorn (\cite{bland95:Moiseev_n}), and a somewhat more detailed
description can be found in the paper by Jones et al. (\cite{jones02:Moiseev_n}). Putting IFP235 into the beam produces ghosts so
that the main image and these ghosts are aligned symmetrically
with respect to the optical axis of the FPI. The intensity of
these ghost reflections (which we refer to  as ``diametral
ghosts'', ``D'', in accordance with the ghost classification
suggested by Bland-Hawthorn (\cite{bland95:Moiseev_n}) and Jones et
al., \cite{jones02:Moiseev_n}) amounts to about 10\% of the intensity of the
main image. In such cases it is usually recommended to tilt the
FPI with respect to the optical axis of the system so as to move
the ghost outside the detector field. However, certain design
features of SCORPIO focal reducer prevent the use of this option.
Therefore during observations the image of the object should be
placed off the optical axis, and care must be taken to ensure
that ghosts from field stars do not fall on the object of interest. All
these problems decrease the working field of view of SCORPIO
focal reducer by a factor of two.

\begin{figure*}[tbp]
\includegraphics[scale=0.9]{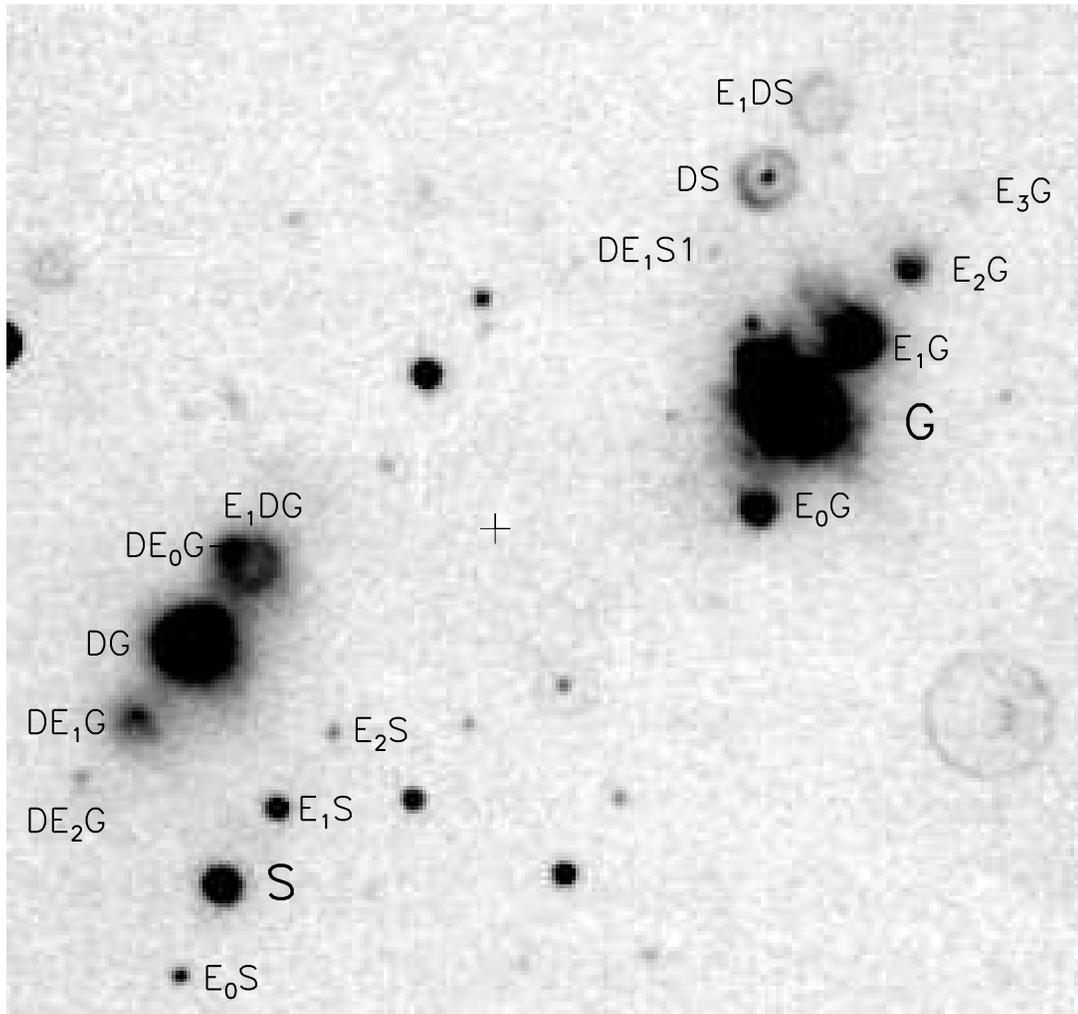}
\caption{Examples of ghost images produced in the IFP501 interferometer. H$\alpha$
observations were performed on January 30/31, 2007 at the request of
C.~Mu\~{n}oz-Tu\~{n}\'{o}n. Here ``G'' and ``S'' denote the II~Zw~70 galaxy and a
bright foreground star, respectively. The sum of all channels in the
data cube is shown. Diametral and exponential ghost images are denoted by ``D'' and
``E'', respectively. The cross indicates the center o the field of view.}
\label{fig1:Moiseev_n}
\end{figure*}

In observations made with IFP501 another---``exponential''
(``E'', according to the terminology of
Bland-Hawthorn (\cite{bland95:Moiseev_n}) and Jones et al., \cite{jones02:Moiseev_n})---family
 of ghost images appears. These ghost images form as a
result of back reflections inside the plates of the FPI and pose
a more serious problem than diametral ghosts, because the nearest
exponential ghost is located within only $16''$ of the object
image, and its intensity is rather high (about 12\% of the
brightness of the object). Such intense ghost images are due to
the degradation of the antireflective coating of the outer
surfaces of the IFP501 plates. Unfortunately, the cost of plate
replacement (or recoating) in this case is close to that of
assembling a new FPI, which is rather high. We nevertheless were
tempted to use IFP501 before a new interferometer was acquired,
because the former performs well in terms of other characteristics---first
and foremost, because of its  $R\approx8300$ spectral resolution at
the H$_\alpha$ line wavelength, which is rather high for extragalactic
astronomy. Such an instrument proved to be very popular among the observers at
the 6-m telescope for tasks that are impossible to perform with IFP235 because
of its lower spectral resolution. These were primarily observations of nebulae
with relatively small (within 100--200 km/s) range of radial velocities:
Herbig--Haro objects, star-forming regions in the Milky Way and other galaxies,
etc.

In a number of cases ghost traces can be removed at the stage of data reduction by applying
the appropriate algorithms. Figure~\ref{fig1:Moiseev_n} (whose idea we borrowed from
Bland-Hawthorn (\cite{bland95:Moiseev_n}) and Jones~et~al., \cite{jones02:Moiseev_n}) clearly
illustrates the complex pattern of  ghost images.
Namely, the image of the galaxy (G) produces closely-spaced ghost images E$_0$G
and E$_1$G
whose brightness is equal to  4\% and 12\% of the object brightness, respectively. Now,
E$_1$G produces E$_2$G, which, in turn, produces E$_3$G---and the brightness of these
ghosts is equal to 1.4\% and 0.15\% of the brightness of G, i.e., the brightness of ghosts
decreases exponentially as it must be in the case of multiple reflections. These ghosts appear at
the front plate of the FPI before light passes through the interferometer, and
therefore are referred to as monochromatic---their relative intensity in the
data cube does not change with wavelength $\lambda$. At the same time, ghost
E$_0$G appears at the back plate of the  FPI after interference has taken place.
The image of such a ghost on the interferogram  is located in a domain
corresponding to other wavelengths. Hence the data cube should exhibit an offset
in $\lambda$ between the spectra of G and E$_0$G, and it must depend on the
position of the image on the detector.

In Fig.~\ref{fig1:Moiseev_n} we also indicate diametral ghosts, i.e., ghosts located
symmetrically with respect to the optical axis of the interferometer. The image
of galaxy G and the corresponding ghost DG are located symmetrically with
respect to the optical axis, ghost E$_1$G produces diametral ghost DE$_1$G, and
ghost E$_2$G, respectively, produces DE$_2$G, etc. Ghost DG also generates a
family of secondary exponential ghosts: E$_1$DG can be seen in the figure.
The latter is superposed by diametral ghost DE$_0$G. Diametral ghosts have
rather complex structure, as is immediately apparent in the case of ghost DS
produced by the brightest foreground  star. Namely, this ghost consists of two
images of the star---a normal image (with the intensity equal to $\sim5\%$ of
that of star S) and a strongly defocused image, which has the shape of a ring,
whose total intensity is equal to about 10\% of intensity S.

Note that it is by no means always possible to find so many ghost images as in
 Fig.~\ref{fig1:Moiseev_n}.
The figure illustrates a limit case corresponding to observations
of a relatively bright object, whereas usually only the brightest ghosts (DG,
E$_0$G, and E$_1$G) can be seen. We describe the
pattern of ghost images in such detail, because we believe that our analysis
will help the users better understand the data obtained and distinguish real
emission features from ghosts.

If we denote the ideal data cube (i.e., the cube without ghosts) as $I_{real}$, then the
wavelength-calib\-rated observed cube has the following form:

\begin{equation}
I_{obs}(x,y,\lambda)=I_{real}(x,y,\lambda)+I_{ghost}(x,y,\lambda),
\label{eq1:Moiseev_n}
\end{equation}

\noindent where the family of exponential ghosts can be represented in the form:

\begin{equation}
\begin{array}{rr}
I_{ghost}(x,y,\lambda)\approx f_0\hat{P}\,I_{real}^{-1}(x-\Delta x_0,y-\Delta y_0,z)+  \\
+\sum_{i=1}^{i_{max}}f_iI_{real}(x-\Delta x_i,y-\Delta y_i,\lambda).\\
\end{array}
\label{eq2:Moiseev_n}
\end{equation}

Relative brightness $f_i$ of ghosts decreases exponentially with the number $i$ of reflections
and therefore summation is performed not to infinity, but up to $i_{max}=3-5$, because the
contribution of higher-order ghosts is usually insignificant. For the same reason, we neglect
the contribution of secondary ghosts, such as non-monochromatic satellites of the ghosts of the
monochromatic family. The first term in (\ref{eq2:Moiseev_n}) describes the wavelength-shifted
 ghost
E$_0$G. Here $\hat{P}$ denotes  ``phase correction'', i.e., the transformation
from spectra expressed in terms of the FPI channel numbers $z$ to spectra expressed in
terms of the wavelengths:
$$
I(x,y,\lambda)=\hat{P}I(x,y,z)=I(x,y,k(z+p(x,y))),
$$
where $p(x,y)$ is the so-called phase map (see Section~2 in Paper~I for
details), and
$k(z)=k_1z+k_0$ is a linear function whose coefficients are determined by the particular FPI
employed. Correspondingly, $I_{real}^{-1}$ is a result of the reverse transformation from
 wavelength
scale to the scale of interferometer channels:
$$
I_{real}^{-1}(x,y,z)=\hat{P}^{-1}I_{real}(x,y,\lambda).
$$

Our aim is to infer from $I_{obs}$ the best approximation to ghost-cleaned cube $I_{real}$. We
use the following iterative procedure for this. We substitute $I_{obs}$ for
$I_{real}$ in formula (\ref{eq2:Moiseev_n}) to obtain the first approximation to the ghost model
$I_{ghost}^1 (I_{real})$. In this case, the first approximation to the ghost-cleaned cube has,
according to formula (\ref{eq1:Moiseev_n}), the following form:
$I_{real}^1=I_{obs}-I_{ghost}^{1}$. We now substitute  $I_{real}^1$ into formula
 (\ref{eq2:Moiseev_n})
to derive the following approximation to ghost model $I_{ghost}^2$, etc., up to $I_{ghost}^n$.
Because of the small relative brightness of ghost images  ($f_i\ll 1$), only a few iterations
($n=$3--4) are needed to construct a close-to-reality ghost-image model.

The following relations are true for the family of monochromatic ghost images \mbox($i\geq1$):
$$
\Delta x_i=i\Delta x_1,\,\,\Delta y_i=i\Delta y_1,\,\,f_i=f^i,
$$
and hence model (\ref{eq2:Moiseev_n}) has only six free parameters, which are chosen in a way
to minimize the
differences between $I_{ghost}^n$ and $I_{obs}$. The following table gives the mean values
 of these parameters, which vary little from night to night.

\begin{figure*}[tbp] 
\includegraphics[width=16.5cm]{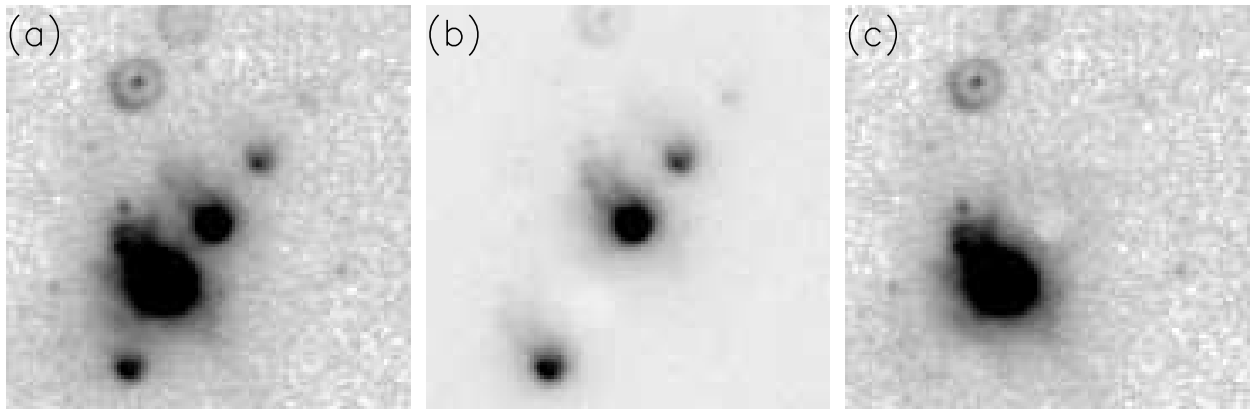}
\caption{Subtraction of ghost images. Here we show the sum of all channels of
the data cube for the II~Zw~70 galaxy: (a) initial image, (b) model of
brightness distribution in ghost images, and (c) the result of subtracting the
model from the initial cube.}
  \label{fig2:Moiseev_n}
\end{figure*}

\begin{table} 
\caption
{Parameters of ghost-image model in H$\alpha$ line}
\label{tab1:Moiseev_n}
\begin{tabular}{c|c|c}
\hline
                    & $i=0$ &  i=1 \\
\hline
$f_i$              &  0.038 & 0.11 \\
$\Delta x_i$, $''$ &  -7.67 & 9.80 \\
$\Delta y_i$, $''$ &  -16.42& 13.63 \\
\hline
\end{tabular}
\end{table}

Our experience in the reduction of the data obtained with IFP501 shows that the model based
on the above algorithm usually describes real ghost images rather well, with no
appreciable bias. The only exception are the data obtained under unstable
atmospheric conditions, when counts in individual channels had to be averaged in
order to compensate for appreciable (more than 10--20\%) variations of the FWHM
of stellar images (for a description of photometric correction see Paper~I.

Figure~\ref{fig2:Moiseev_n} shows how the ghosts around the image of the dwarf galaxy
II~Zw~70 are subtracted. Other examples of observational data processed using
the above de-ghostification procedure can be found, e.g., in
Lozinskaya et al. (\cite{loz_1613:Moiseev_n},\cite{loz_zw:Moiseev_n})  -- a study of star-forming regions in the dwarf galaxies IC1613 and VII Zw403,    Mart\'{i}nez-Delgado et al. (\cite{delgado:Moiseev_n}) -- kinematics of ionized
gas in blue compact galaxies, Movsessian et al. (\cite{movsesian:Moiseev_n}) -- study of outflows from young stellar objects in the HL/HX Tau region.

\begin{figure*}[tbp]
\includegraphics[scale=1.4]{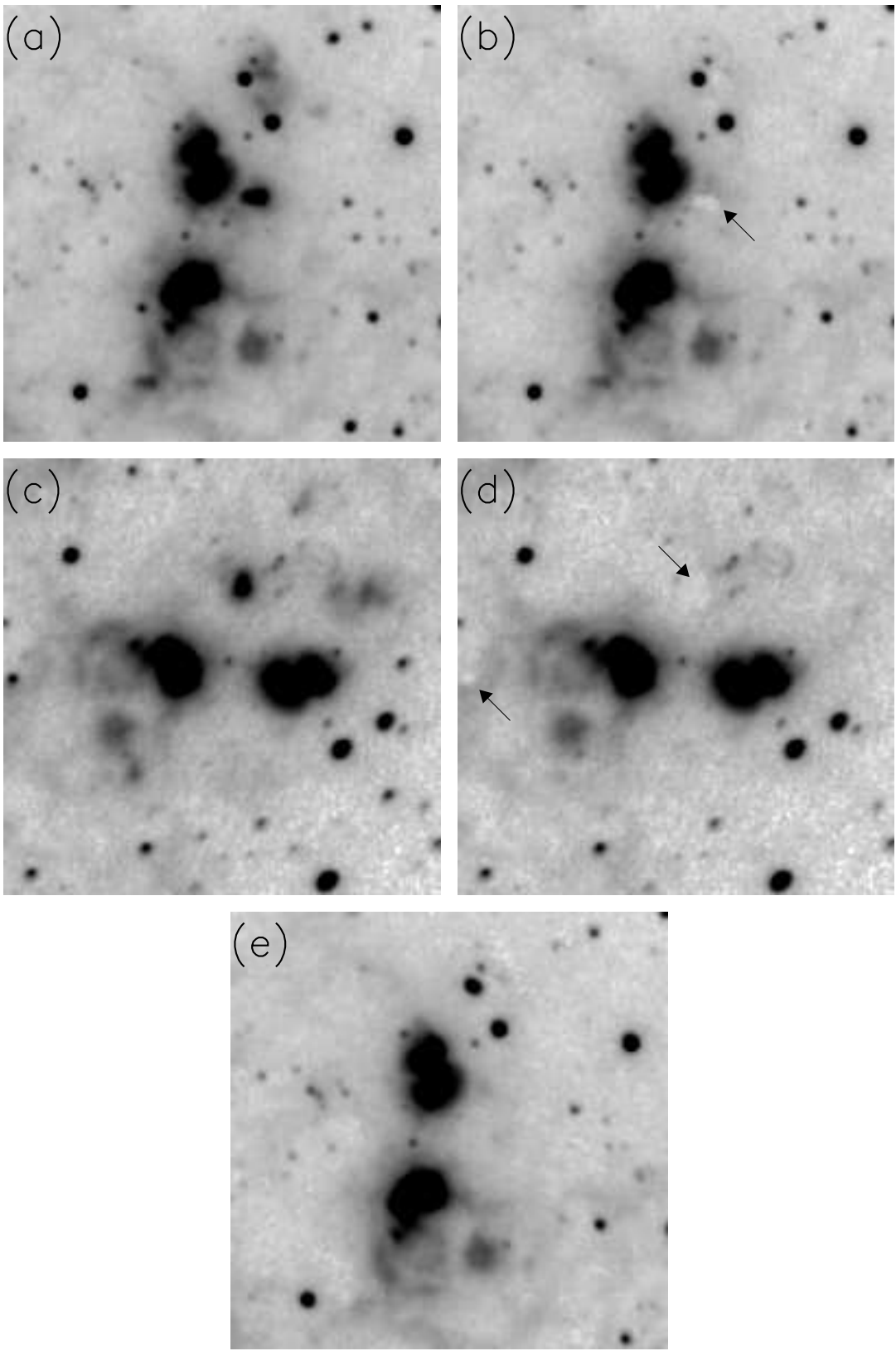}
\caption{Sum of channels in the data cubes for the IC 10 galaxy.
H$\alpha$-line observations were made on September 8/9, 2005 at the request of
T.~A.~Lozinskaya. The figure shows a $72''\times72''$ fragment of the image.
 (a) Initial image in the first cube; (b)  the image after ghost subtraction,
the arrow indicates regions with unsatisfactory quality of subtraction;  (c)
the image in the cube obtained by turning the field of view by $90\degr$;
(d) subtraction of ghosts from the second cube (the arrows indicate
low-quality regions), and (e) the combinations of the two data cubes.} \label{fig3:Moiseev_n}
\end{figure*}

Yet another problem arises in observations of objects with strong
surface-brightness gradients, where an intense ghost of, e.g., the galaxy
nucleus, projects onto regions of much lower brightness. The level of Poisson
noise is determined by the combined intensity of the ghost and the base.
Therefore after the subtraction of the ghost model a situation may arise where
the useful signal in the region considered is comparable to the amplitude of
photon noise. To avoid loss of spectral information from low-brightness regions
in observations of such objects, we recommend to divide the planned
exposure into two and perform observations successively with two different
orientations of the instrument's field of view turned by about $90^\circ$ in
position angle. The ghosts obtained in the two data cubes should then
project onto different regions of the object. After primary reduction, the
corresponding ghost model is subtracted from each data cube and the regions are
masked where the signal-to-noise ratio decreased strongly after ghost removal.
The two cubes are then combined into one and spectra of masked regions in each
cube are substituted by the corresponding  ``good'' spectra from
the other data set. One of the best examples of application of the above
algorithm is the reduction of observations of the nearby dwarf galaxy IC~10
reported by Lozinskaya et al. (\cite{loz_10:Moiseev_n}). Here the image of emission
shells of ionized gas occupies more than half of the entire field of
view of SCORPIO instrument. Figure~\ref{fig3:Moiseev_n}  illustrates the
sequence of operations in the process of subtracting ghosts from the
images of these galaxy.

\section{Velocity dispersion measurement}
\label{disp:Moiseev_n}

A number of observational programs performed with the FPI (e.g., mapping the ionized-gas
velocity dispersion in galaxies) require accurate estimates of the halfwidths
of emission lines. These estimates must take into account the broadening
due to  the instrumental profile. It is common practice to use the
following formula (which assumes that both the instrumental
profile of the spectrograph and the initial---unbroadened---line profile can be described
 by Gaussian functions
with the dispersions equal to  $\sigma_{real}$ and $\sigma_{ins}$,
respectively):

\begin{equation}
\sigma_{obs}=\sqrt{\sigma_{real}^2+\sigma_{ins}^2}.
 \label{eq3:Moiseev_n}
\end{equation}

\noindent Here $\sigma_{obs}$ denotes the dispersion of the Gaussian used to
describe the profile of the line observed at the output of the
instrument. Hereafter by measuring the velocity dispersion we mean  estimating
$\sigma_{real}$ from the observed spectra. One should keep in mind that
broadening of lines in the spectrum of the observed object may be caused not
only by velocity dispersion $\sigma_{gas}$ (the measure of chaotic motions along
the line of sight), but also by a number of other factors. Thus, according to
Rozas et al. (\cite{rozas00:Moiseev_n}), the following formula can be written for the
integrated spectrum of HII regions:

\begin{equation}
\sigma_{real}^2=\sigma_{gas}^2+\sigma_{N}^2+\sigma_{tr}^2,
\label{eq4:Moiseev_n}
\end{equation}

\noindent where $\sigma_{N}^2\approx3$~km/s and
$\sigma_{tr}^2\approx9.1$~km/s correspond to the natural width of the emission
line and its  thermal broadening at  $10^4$ K, respectively.

Because of its simple and self-explanatory form, formula (\ref{eq3:Moiseev_n}) is
widely used to analyze spectroscopic data. It is often generalized by
substituting FWHM for $\sigma$. However, one must keep in mind that
exact equality in (\ref{eq3:Moiseev_n}) is achieved only for Gaussian functions,
because a convolution of two Gaussians is also a Gaussian. The assumption
about the Gaussian form of the instrumental profile is usually true for slit
spectrographs. However, the instrumental profile of the FPI, which is
given by the Airy function, has wide wings and is best approximated by
a Lorentz profile (see, e.g., Bland-Hawthorn (\cite{bland95:Moiseev_n}) and Paper~I) rather than by a Gaussian. Therefore if the initial profile of the emission line is a Gaussian
with the dispersion determined by formula (\ref{eq4:Moiseev_n}), then the observed
profile is the convolution of the Gaussian and Lorentz profiles and is hence
given by the Voigt function:

\begin{equation}
V(\lambda,y)=\frac{1}{\sqrt{2\pi}\sigma_{real}}\frac{y}{\pi}\int_{-\infty}^{\infty}\frac{e^{-x^2}dx}{y^2+(a-x)^2},
\label{eq5:Moiseev_n}
\end{equation}

\noindent where

$$
a=\frac{\lambda-\lambda_0}{\sqrt{2}\sigma_{real}},~~~
y=\frac{w_{ins}}{\sqrt{2}\sigma_{real}}.
$$

\begin{figure}[tbp]
\includegraphics[scale=0.8]{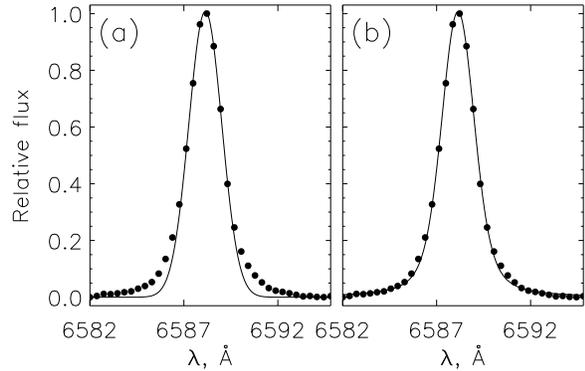}
\caption{Example of the  H$\alpha$-line spectral profile in the
II~Zw~70 galaxy based on observations with IFP501 (dots). The solid line shows
Gaussian (a) and Voigt  (b) fits.}
 \label{fig4:Moiseev_n}
\end{figure}

\noindent Here $\lambda_0$ is the central wavelength and $w_{ins}$ denotes the
halfwidth of the instrumental (Lorentz) profile of the FPI as
determined from the spectrum of the lines of the calibration lamp. We then
approximate the observed profile by function (\ref{eq5:Moiseev_n}) to obtain the required
estimate $\sigma_{real}$. It is evident from Fig.~\ref{fig4:Moiseev_n} that compared to
the Gaussian the Voigt profile fits much better the line wings in the
galaxy spectra observed with IFP501. It is clear from general considerations
that formula (\ref{eq5:Moiseev_n}) is a more correct tool for estimating the velocity
dispersion than formula (\ref{eq3:Moiseev_n}). In the latter case we have to use Gaussian
approximation for the profiles of the lines that deviate systematically from
the adopted approximation (Fig.~\ref{fig4:Moiseev_n}a), and this may introduce
an additional error in the estimated  $\sigma_{real}$. However, this approach
is highly popular in velocity-dispersion measurements for extragalactic HII
regions (see, e.g.,  Mart\'{i}nez-Delgado, \cite{delgado:Moiseev_n};  Mu\~{n}oz-Tu\~{n}\'{o}n, \cite{munoz95:Moiseev_n}; Rozas et al. \cite{rozas00:Moiseev_n}). We believe
that this method owes its popularity  not only to the less
complex appearance of formula (\ref{eq3:Moiseev_n}) compared to that of formula
(\ref{eq5:Moiseev_n}), but also to the fact that Gaussian approximation of spectral-line
profiles is incorporated into virtually all packages of astronomical data
reduction. The use of the Voigt profile for analyzing extragalactic
spectra is less common (see, e.g., \cite{gebh94:Moiseev_n}), despite the  fact that
integration in  (\ref{eq5:Moiseev_n}) poses no problem for modern computers.

\begin{figure*}[tbp]
\includegraphics[scale=.9]{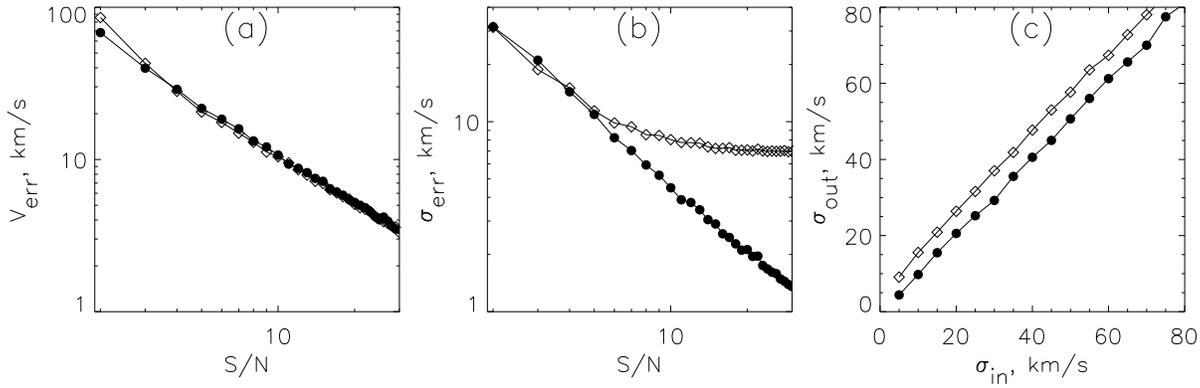}
\caption{Simulation of errors of measurement of kinematic
parameters (at the $1\sigma$ level). The computations were  made for IFP501.
The diamond signs and filled circles show the Gaussian and Voigt-profile
approximations, respectively. (a) Dependence of the error of
radial-velocity measurements on the signal-to-noise ratio. The average velocity
dispersion was equal to  50 km/s; (b) error of measured velocity
dispersion as a function of signal-to-noise ratio under the same
conditions; (c) comparison of the initial and measured velocity
dispersion (for $S/N=20$).}
\label{fig5:Moiseev_n}
\end{figure*}

\begin{figure*}[tbp]
\includegraphics[scale=.9]{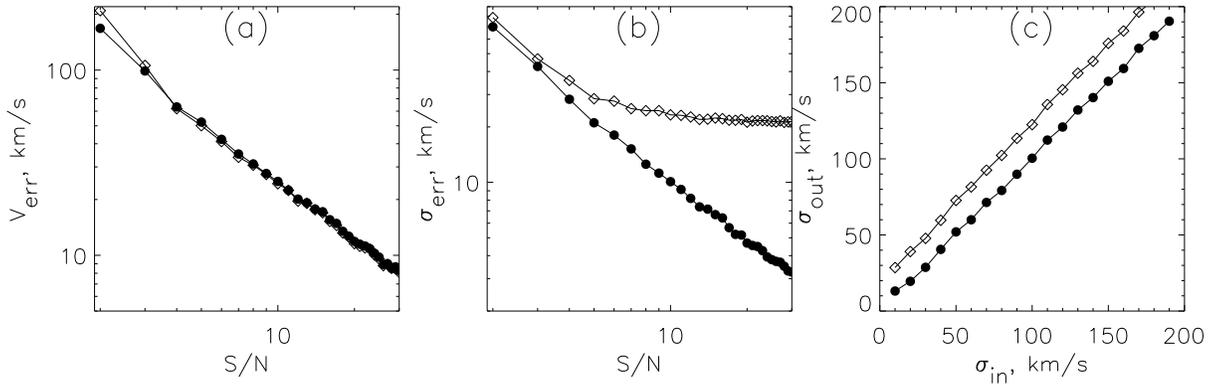}
\caption{\protect Same as Fig.~\ref{fig5:Moiseev_n}, but for IFP235. Figures~(a) and
(b) are for the case where an average velocity dispersion is equal to 200~km/s. }
\label{fig6:Moiseev_n}
\end{figure*}

We estimated the errors of measurement of the kinematical parameters
(radial velocity and velocity dispersion) for both approaches considered.
We smoothed the instrumental profile of the FPI by a Gaussian with
the dispersion equal to $\sigma_{in}$ and then added noise to the resulting
spectrum and estimated the velocity dispersion using both methods. The
difference between the output ($\sigma_{out}$) and input ($\sigma_{in}$)
velocity dispersions allows us to estimate the error  $\sigma_{err}$ of the
velocity dispersion measuring. We obtained a total of 1000 independent measurements for
each fixed signal-to-noise ratio ($S/N$). We similarly estimated the
error of measured radial velocity. We performed our computations for
the instrumental profiles with the width near the H$\alpha$ line equal to
$w_{ins}=35$ and $w_{ins}=115$ km/s for IFP501 and IFP235, respectively.
Figures~\ref{fig5:Moiseev_n}\,(a,\,b) and \ref{fig6:Moiseev_n}\,(a,\,b) show the results of
 computations---radial-velocity and velocity-dispersion errors as functions of the signal
level. As expected, the  error of radial-velocity measurements for
symmetric lines does not depend on the algorithm employed. At the
$S/N=30$ it is equal to 2.5 and 8 km/s  for interferometers IFP501
and IFP235, respectively.

The situation is quite different for the error of the measured velocity
dispersion. As expected, the error of measurements based on Voigt-profile
fits smoothly decreases with decreasing signal level and amounts to only
several km/s for $S/N>20$. These errors are due to noise in the measured
spectra and they contain no systematic component. Contrariwise, if
measurements are based on relation (\ref{eq3:Moiseev_n}) then systematic but not
random component of measurement error starts to dominate at $S/N\geq10$. More
specifically,  velocity dispersion is overestimated, as it is evident from
Figs. ~\ref{fig5:Moiseev_n}c and \ref{fig6:Moiseev_n}c. Velocity-dispersion estimates inferred
from Gaussian fits to the profiles exceed the actual values by 7--8 and
20--25 km/s for IFP501 and IFP235, respectively. Such a systematic error
is unimportant for estimating widths of lines with $\sigma_{real}>100$ km/s.
However, velocity dispersion may be overestimated by up to 100\% in the case
$\sigma_{real}=10-20$  km/s. This may be of critical importance, e.g., in the
studies of ionized gas in star-forming regions, when it is necessary to identify
expanding shells or regions where the velocities of chaotic motions exceed the
speed of sound in the interstellar medium  (Mart\'{i}nez-Delgado et al. \cite{delgado:Moiseev_n}; Rela\~{n}o \&  Beckman, \cite{relano05:Moiseev_n}).
 In these cases
velocity dispersion estimates inferred by fitting the Voigt profile are to be
used. Or, if for some reasons the authors prefer Gaussian approximation, it is
necessary to estimate systematic errors like we did it above and correct
correspondingly the estimates based on formula (\ref{eq3:Moiseev_n}).

Note that Rela\~{n}o \&  Beckman (\cite{relano05:Moiseev_n}) proposed an alternative
method to account for the instrumental profile of the FPI using
the reconstruction (deconvolution) technique. This procedure is to be used
for multicomponent lines. However, the method can be applied only to
spectra with sufficiently high signal-to-noise ratios.

\begin{figure*}[tbp]
\includegraphics[scale=0.9]{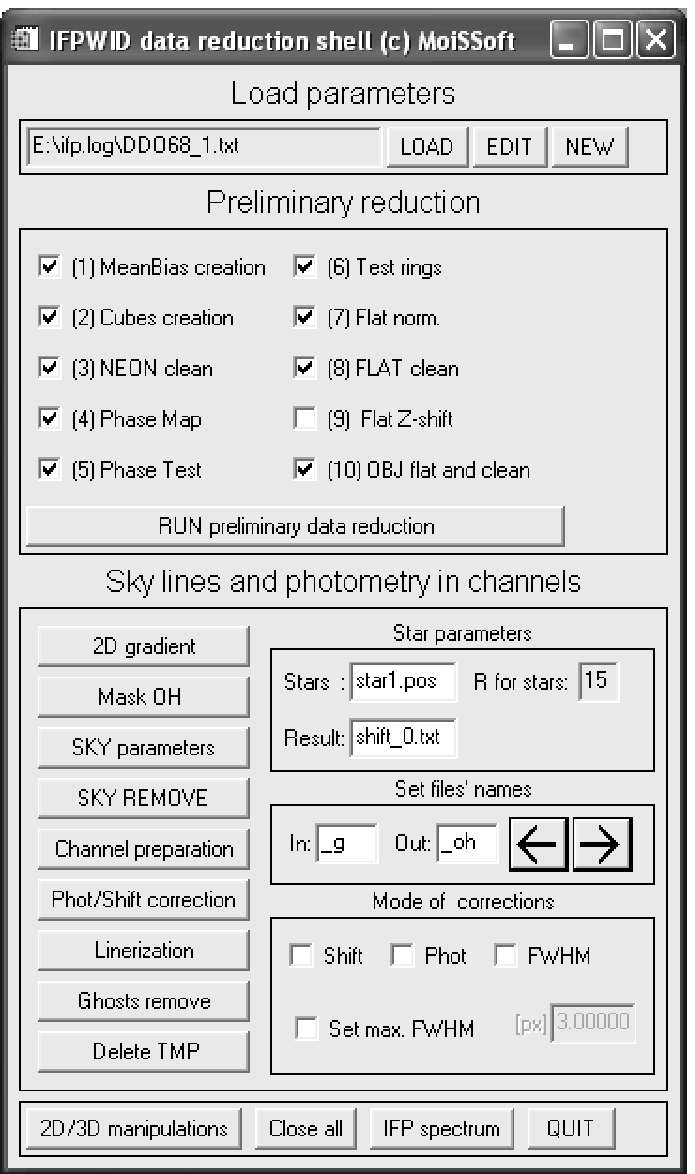}
\includegraphics[scale=0.9]{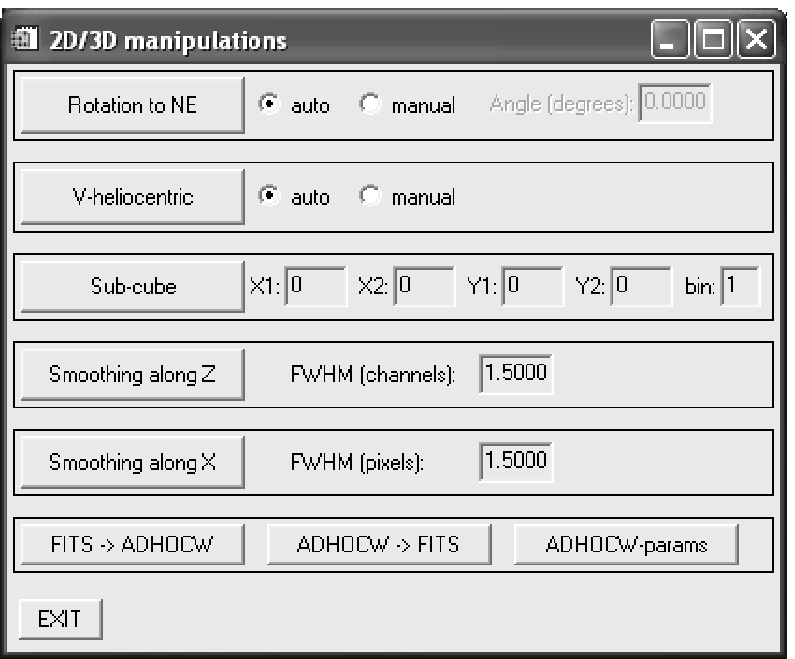}
\caption{IFPWID interface: the main menu (on the left) and the menu of operations
with calibrated cubes (on the right).} \label{fig7:Moiseev_n}
\end{figure*}

\section{Brief description of the software package}
\label{soft:Moiseev_n}

To reduce observations made with the FPI on SCORPIO, we wrote IFPWID
program package with multiwindow user-friendly interface in IDL 6.X language
(Fig.~\ref{fig7:Moiseev_n}).~~ The~~ codes~~ are~~ publicly~~ available~~ at
{\tt http://www.sao.ru/hq/moisav/soft}. Below in this section we describe the
principal sequence of data-reduction steps to be made when using these programs.
We do not describe in detail most of the algorithms employed---wavelength-scale
calibration, photometric correction, night-sky line
subtraction, etc. (see our sufficiently detailed Paper~I, and only
briefly list the initial data for the reduction of a set of CCD frames:

\begin{itemize}

\item \textit{OBJECT}---inteferograms of the object studied.

\item \textit{NEON}---images of interference rings from the
emission line selected by the narrow-band filter from the spectrum of
the He-Ne-Ar calibration lamp. This calibration  is usually
performed before and after the observing night.

\item \textit{FLAT}---interferograms of uniform ``flat-field'' illumination
produced by the continuous-spectrum lamp and obtained with the same narrow-band
filter as was used for observations of the object.

\item \textit{TEST}---images of the rings from the line-spectrum
lamp in some FPI channels obtained along with object integrations at the
same position of the telescope. These images are used to monitor the accuracy
of scanning and to control the offset of the center of the rings due to the instrument
flexures.

\end{itemize}

\subsection{Assembling the Data Cubes}

The first stage of reduction usually requires no special
settings. As a rule, it suffices to indicate only the number of
the observational night, file-name templates, and the required
frame format (some observations with the FPI involve the readout
of only a fragment of the CCD). The program extracts all the
remaining necessary information (parameters of the FPI, numbers
of spectral channels, etc.) from the descriptors of the corresponding FITS
files, which are filled automatically in the process of observations.
Therefore the user only has to check the required reduction steps in the
menu (see the upper part of the menu in Fig.~\ref{fig7:Moiseev_n}) and press the ``run''
button. Below we briefly list these steps (their names used in the
reduction program are italicized):

\begin{itemize}

\item Search for and averaging of the bias-current frames taken in the
required mode of CCD readout (\textit{MeanBias creation}). The
resulting superbias frame is then subtracted from all object integration and
calibration frames. The CCD employed contains virtually no ``hot'' pixels,
dark current is small and therefore can be neglected for exposures about
several minutes.

\item Creation of data cubes from individual frames (\textit{Cubes creation}).
Superbias is subtracted and bad columns are masked. Further operations are
the reduction of three-dimensional cubes (object interferograms
and calibration data assembled in order of channels). Cubes are
stored in the standard  FITS format (NAXIS=3).

\item Removal of cosmic-ray hits from the cubes of calibrating-lamp integration
with line (\textit{NEON clean}) and continuous (\textit{FLAT clean}) spectra.
Simple $\sigma$-filter is used here: the counts in the spectra that
deviate from the mean by more than preset threshold value are substituted by the
half-sum of the neighboring channels.

\item Construction of the phase-shift map based on the results of
Lorentz-profile fits to the lines in the NEON cube (\textit{Phase
map}). A correction is applied, where necessary, to allow for nonuniform
scanning of the calibrating cube (see Paper~I).

\item Testing the accuracy of wavelength scale (\textit{Phase test}). The
NEON cube is corrected for the phase shift and then the position
of the emission line of the calibration lamp is measured for each
pixel.

\item Computation of corrections to the wavelength scale constructed
from the calibration cube  (\textit{Test rings}). The offsets (along both
coordinates in the CCD plane and along the wavelength coordinate) of TEST
frames are computed relative to the NEON cube (to correct for
the instrument flexures and to monitor the scanning accuracy of the FPI).

\item Correction (if needed) of relative variations of FLAT lamp
brightness during scanning of the calibration cube (\textit{Flat
norm}).

\item Measurement of the offsets between the transmission maxima of the
narrow-band filter measured for the FLAT cube and for the spectra of selected
 stars in the field of the object (\textit{Flat Z-shift}). This procedure
is usually needed only for observations made with IFP235, where the
half-width of the narrow-band filter is appreciably smaller than the
wavelength interval between the neighboring orders of interference. The offset
is usually close to zero, however, it may reach $2-3.5$\,\AA~for some
filters.

\item Division of the cube by the flat-field (to correct for the transmission
curve of the narrow-band filter) and cosmic-ray hit removal (\textit{OBJ flat
and clean}).

\end{itemize}

\subsection{Subtraction of Night-Sky Lines and Conversion to the Wavelength Scale}

Then  comes the turn of the reduction procedures that require
the user's intervention  more often than during the stage of
assembling the data cubes. The procedures correspond to the
buttons of the main interface  (the left-hand part of
Fig.~\ref{fig7:Moiseev_n}a), which are italicized below. Reduction is
usually performed in the following order:

\begin{itemize}

\item Correction for the background gradient (\textit{2D gradient}).
In case of observations
of objects in some narrow-band filters the background brightness
distribution in the object cube exhibits appreciable gradient
even after the division by the FLAT, especially in the presence
of additional light pollution due to the Moon. This residual
gradient must be due to nonuniform illumination from the  FLAT
lamp combined with specifics of the interference coatings of
particular filters. The user can correct this effect by setting
the parameters of the two-dimensional brightness distribution to
which the object cube is normalized.

\item Creation of the mask for subtracting the spectrum of the night sky (\textit{Mask OH}).
The mask is based on the image of the sum of the channels of the
object cube. Regions with the brightness below the given
threshold are considered as ``background''. The resulting mask
can be edited if necessary.

\item Setting the parameters for the sky-background subtraction (\textit{SKY parameters}).
We described our technique of sky-background subtraction in the Paper~I. Its main idea is to average sky background emission in each channel of the  OBJ cube over the
azimuthal angle within narrow rings centered on the optical axis
of the FPI with taking into account the mask constructed at the
previous stage of reduction. The average brightness profile is
then subtracted from the interferogram of the object. The user
may choose various modes of such subtraction, vary the width of
the rings where averaging is performed, fix the center of the
rings or set options for an automatic search of the ring center in each
channel. When needed, averaging can be performed within individual
sectors (such a procedure can be useful for correcting for variations
of the instrumental contour over the field of view), etc.

\begin{figure*}[tbp]
\includegraphics[width=15cm]{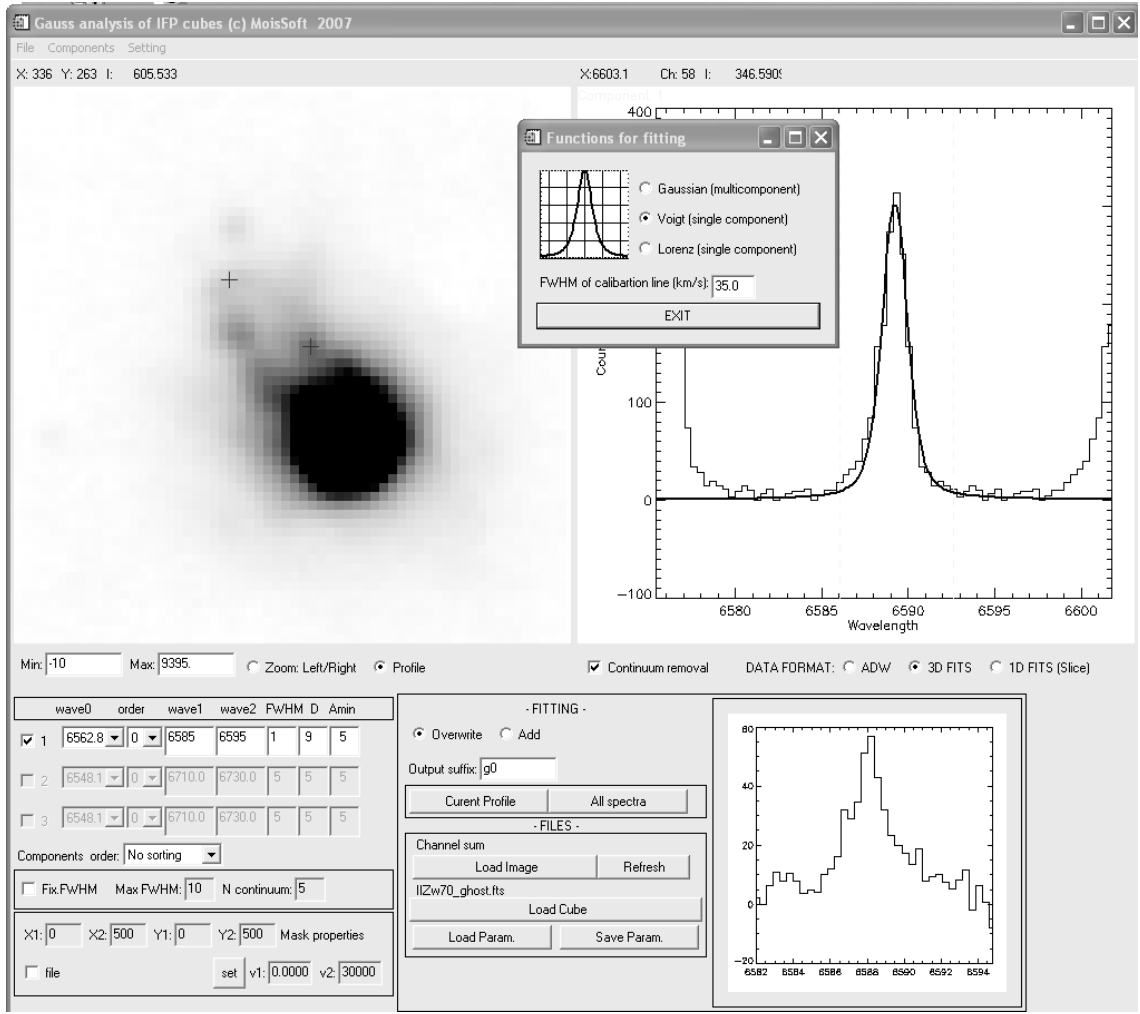}
\caption{Interface of the program of the analysis of
emission-line profiles.} \label{fig8:Moiseev_n}
\end{figure*}

\item Subtraction of the night-sky spectrum in accordance with the
parameters given above (\textit{Sky remove}).

\item Photometry of stars in each channel of the object cube (\textit{Channels preparation}).
The relative offsets of the image centers, variations of FWHM and
integral flux are measured for stars from a precompiled list and
the averaged dependences of these parameters on the number of the
channel in the cube are constructed.

\item Photometric correction of the object cube based on the
results of photometry of stars in each individual channel
(\textit{Photo/Shift corrections}). Corrective channel offsets
are applied and account is taken of variations of atmospheric
transparency (the channel counts are multiplied by the computed
coefficients) and seeing (convolution with two-dimensional
Gaussians).

\item Conversion of the object cube to the wavelength scale (\textit{Linearization}).

\item Subtraction of ghosts using the procedure described in
Section \ref{ghost:Moiseev_n}
 (\textit{Ghosts remove}).
\end{itemize}

\subsection{Processing of Calibrated Cubes}

The reduction steps described above produce a wavelength-calibrated data
cube  with maximum account taken of all instrumental effects. In
this form it can be analyzed using various software tools
depending on the user preferences or on the task to be
accomplished. The data can also be converted to the format
adopted in the popular ADHOC\footnote{ADHOC software package was
developed by J.~Boulesteix (Observatoire de Marseille) and is
available at \texttt{http://www.oamp.fr/adhoc/adhocw.htm}}
reduction system. Below we describe the operating sequence to be
performed with our software tools in order to map emission-line
radial-velocity and velocity-dispersion distributions.

Figure \ref{fig7:Moiseev_n} (on the right) shows the menu of operations with
calibrated cubes. The following procedures (their names are
italicized) are performed:

\begin{itemize}
\item Rotation of the cube to the ``correct'' orientation of the images (with North
at the top and East on the left), because observations can be
performed at any arbitrary (or specially selected) position angle
(\textit{Rotation to NE}).  The turn angle is computed from the
data of the descriptors of the FITS-file header to within 0.1\degr.   
In the cases where better accuracy is required astrometric
reduction is to be performed using field stars.

\item Correcting the radial velocities for the motion of the Earth
about the Sun (\textit{V-heliocentric}). The necessary
information is extracted from the FITS-file header.

\item If necessary, a cube fragment is cut containing the object studied
(\textit{Sub-cube}). In this form reduced FPI cubes are usually
stored in our  ASPID database \cite{aspid07:Moiseev_n}.

\item Smoothing the cubes using a one- and two-dimensional Gaussians
of given width along the spectral  (\textit{Smoothing along Z)} and
spatial (\textit{Smoothing along X}) coordinates, respectively.
\end{itemize}

We construct the velocity fields, velocity-disper\-sion maps,
monochromatic images in the emission line and in the continuum,
using GAUS program whose main interface is shown in
Fig.~\ref{fig8:Moiseev_n}. This program can be used to analyze individual
profiles in selected pixels of the data cube by fitting the
emission-line profiles to Voigt and Gauss functions. The integral
in  (\ref{eq5:Moiseev_n}) is computed via standard  VOIGT function of the
IDL language. The main task performed by the program
is an automatic identification of emission lines in the
data cube and their approximation by various functions for the
given parameters. This is important because the number of pixels
that contain useful signal in our data cubes may amount to
several hundred thousands, thereby preventing individual approach
to each spectrum. Our procedure yields two-dimensional maps of
profile parameters (Doppler velocities, line fluxes, etc.).

\section{Conclusions}

In this paper we briefly describe the software currently used to
reduce observations made with the scanning FPI operated as a part
of SCORPIO instrument. We believe the software complex described
here to fully meet the requirements imposed by the research tasks
performed at the 6-m telescope of the SAO RAS using the
observational technique considered. In our opinion, further
evolution of the data reduction software package should be
associated with project of the development of SCORPIO-2
new-generation multimode focal reducer at the SAO RAS. We
hope, first and foremost, that the new illuminator of the
calibration beam would make it possible to abandon flat-field
correction procedures (nonuniform residual background, FLAT cube
wavelength shift observed in some narrow-band filters etc.).
Higher automation level of SCORPIO-2 (compared to the current
version of the instrument) will allow further unification of the
process of acquisition data with the FPI and manage
without ``manual'' setting of parameters in a number of
procedures. The latter primarily concerns subtraction of the
night-sky spectrum. In this case it will suffice to simply
develop a pipeline for the reduction of the data obtained with
the scanning FPI.

\begin{acknowledgements}
We are grateful to A.~A.~Smirnova for her assistance in the
preparation of the text. This work was supported by the Russian
Foundation for Basic Research (project no.~06-02-16825) and a
grant of the President of the Russian Federation (project
MK1310.2007.2).
\end{acknowledgements}

\end{document}